\documentclass[conference]{IEEEtran}
\usepackage{graphicx}
\usepackage{amssymb,amsmath}
\usepackage{xcolor}
\usepackage{algorithm}
\usepackage{algpseudocode}
\usepackage{extarrows}
\usepackage{stackengine}
\usepackage{tikz}
\usepackage{textcomp}
\usepackage{siunitx}
\usepackage{soul}

\begin{document}
\onecolumn
\section*{IEEE Copyright Notice}
\textcopyright$\;$2020 IEEE. Personal use of this material is permitted. Permission from IEEE must be obtained for all other uses, in any current or future media, including reprinting/republishing this material for advertising or promotional purposes, creating new collective works, for resale or redistribution to servers or lists, or reuse of any copyrighted component of this work in other works. 

\textbf{Accepted to be published in: Proceedings of the 2020 IEEE International Symposium on Hardware Oriented Security and Trust (HOST), May 4-7, 2020, San Jose, CA, USA.}
\twocolumn
\newpage

\title{MaskedNet: The First Hardware Inference Engine Aiming Power Side-Channel Protection}

\author{
    \IEEEauthorblockN{Anuj Dubey\IEEEauthorrefmark{1}, Rosario Cammarota\IEEEauthorrefmark{2}, Aydin Aysu\IEEEauthorrefmark{1}}
    \IEEEauthorblockA{\IEEEauthorrefmark{1}Department of Electrical and Computer Engineering, North Carolina State University
    \\\{aanujdu, aaysu\}@ncsu.edu}
    \IEEEauthorblockA{\IEEEauthorrefmark{2}Intel AI, Privacy and Security Research
    \\rosario.cammarota@intel.com}
}
\maketitle              % typeset the header of the contribution

\begin{abstract}
Differential Power Analysis (DPA) has been an active area of research for the past two decades to study the attacks for extracting secret information from cryptographic implementations through power measurements and their defenses.
The research on power side-channels have so far predominantly focused on analyzing implementations of ciphers such as AES, DES, RSA, and recently post-quantum cryptography primitives (e.g., lattices).
Meanwhile, machine-learning applications are becoming ubiquitous with several scenarios where the Machine Learning Models are Intellectual Properties requiring confidentiality. 
Expanding side-channel analysis to Machine Learning Model extraction, however, is largely unexplored.

This paper expands the DPA framework to neural-network classifiers. 
First, it shows DPA attacks during inference to extract the secret model parameters such as weights and biases of a neural network.
Second, it proposes the \emph{first countermeasures} against these attacks by augmenting \emph{masking}.
The resulting design uses novel masked components such as masked adder trees for fully-connected layers and masked Rectifier Linear Units for activation functions.
On a SAKURA-X FPGA board, experiments show that the first-order DPA attacks on the unprotected implementation can succeed with only 200 traces and our protection respectively increases the latency and area-cost by 2.8$\times$ and 2.3$\times$.

\begin{IEEEkeywords}
Machine Learning, Neural Networks, Side-Channel Analysis, Masking.
\end{IEEEkeywords}

\end{abstract}

\vspace{-.25em}
\section{Introduction}
Since the seminal work on Differential Power Analysis (DPA)~\cite{kocher99}, there has been an extensive amount of research on power side-channel analysis of cryptographic systems. 
Such research effort typically focus on new ways to break into various implementations of cryptographic algorithms and countermeasures to mitigate attacks. 
While cryptography is obviously an important target driving this research, it is not the only scenario where asset confidentiality is needed---secret keys in the case of cryptographic implementations.

In fact, Machine Learning (ML) is a critical new target with several motivating scenarios to keep the internal ML model secret. The ML models are considered trademark secrets, e.g., in Machine-Learning-as-a-Service applications, due to the difficulty of training ML models and privacy concerns about the information embedded in the model coefficients such as weights and biases in the case of neural networks.
If leaked, the model, including weights, biases and hyper-parameters can violate data privacy and intellectual property rights. Moreover, knowing the ML classifier details makes it more susceptible to adversarial ML attacks~\cite{Dalvi}, and especially to test-time evasion attacks\cite{ML-poisoning1,ML-poisoning2}.
Finally, ML has also been touted to replace cryptographic primitives~\cite{kanter02}---under this scenario, learning the ML classifier details would be equivalent to extracting secrets from cryptographic implementations.

\begin{figure}[t!]
\centering
  \includegraphics[width=0.45\textwidth]{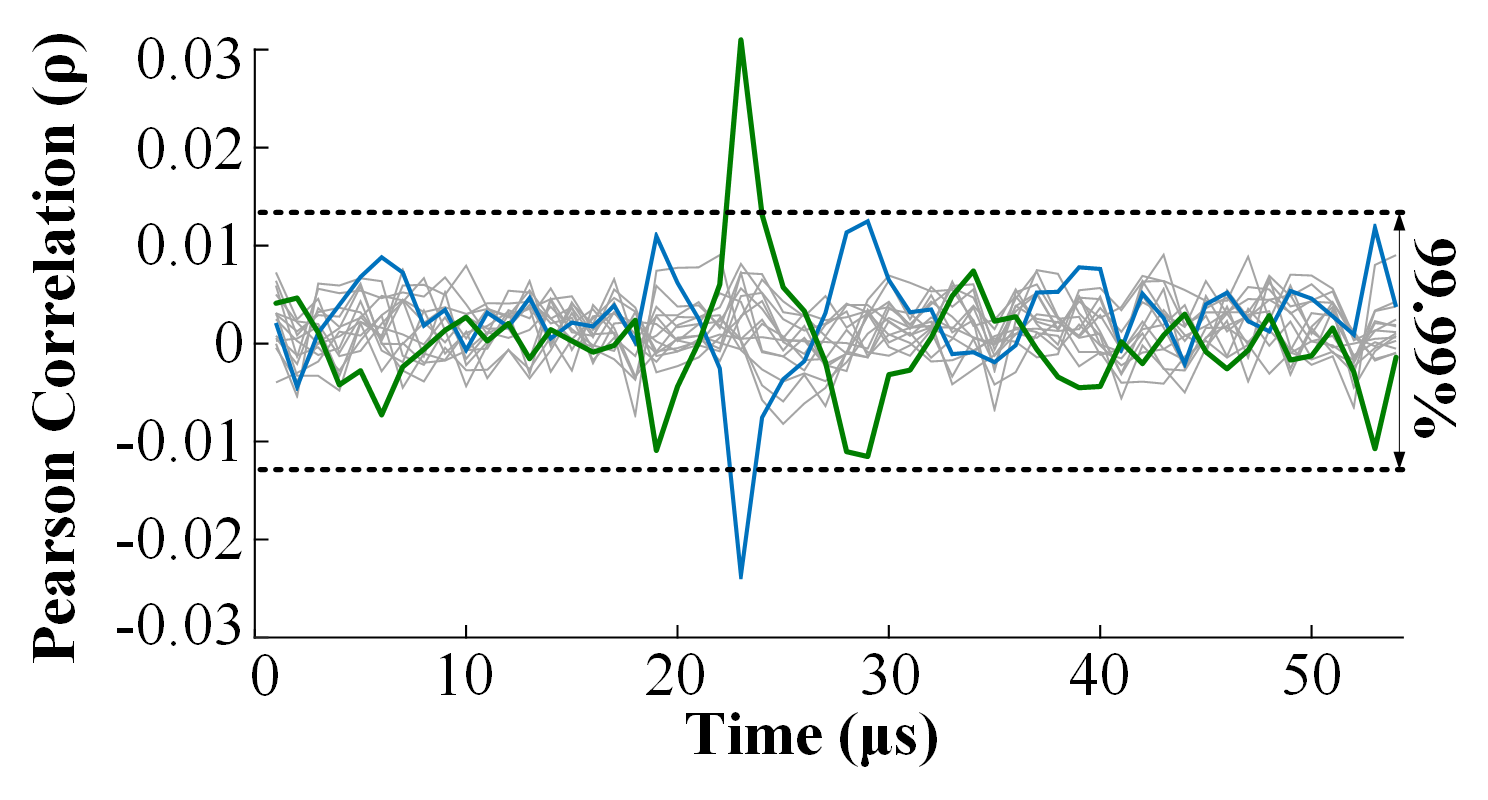}
   \vspace{-1.5em}
   \caption{Motivation of this work: DPA of the BNN hardware with 100k measurements. Green plot is the correlation trace for the correct 4-bit weight guess, which crosses the 99.99\% confidence threshold revealing the significant leak. Blue plot is for the 2's complement of the correct guess, which is an expected false positive of the target signed multiplication. Other 14 guesses do not show correlation.}
   \vspace{-2em}
\label{fig:DPA_attack}
\end{figure}

In this work, we extend the side-channel analysis framework to ML models. Specifically, we apply power-based side-channel attacks on a hardware implementation of a neural network and propose the \emph{first} side-channel countermeasure. Fig. \ref{fig:DPA_attack} shows the vulnerability of a Binarized Neural Network (BNN)---an efficient network for IoT/edge devices with binary weights and activation values~\cite{Courbariaux}. Following the DPA methodology~\cite{brier04}, the adversary makes hypothesis on 4 bits of the secret weight. For all these 16 possible weight values, the adversary computes the corresponding power activity on an intermediate computation, which depends on the known input and the secret weight. This process is repeated multiple times using random, known inputs. The correlation plots between the calculated power activities for the 16 guesses and the obtained power measurements reveal the value of the secret weight.

Fig. \ref{fig:DPA_attack} shows that at the exact time instant where the targeted computation occurs, a significant information leakage exists between the power measurements and the correct key guess. The process can be repeated reusing the same power measurements to extract other weights. Hence, it shows that implementations of ML Intellectual Properties are also susceptible to side-channel attacks like ciphers. 

Given this vulnerability, the objective of this paper is to examine it and propose the first countermeasure attempt for neural network inference against power-based side-channel attacks.
A neural network inference is a sequence of repeated linear and non-linear operations, similar in essence to cryptographic algorithms, but has unique computations such as row-reduction (i.e., weighted summation) operations and activation functions. 
Unlike the attack scenario, the defense exhibits challenges due to the presence of these operations in neural networks, which introduce an additional and subtle type of leak.
To address the vulnerability, we propose a countermeasure using the concepts of message blinding and secret sharing. This countermeasure style is called masking~\mbox{\cite{coron00}}, which is an \emph{algorithm-level} defense that can produce resilient designs independent of the implementation technology~\mbox{\cite{nikova06}}. We tuned this countermeasure for the neural networks in a cost effective way and complement it with other techniques.

\vspace{0.25em}
The main contributions of the paper include the following:
\vspace{-0.25em}
\begin{itemize}
    \item We demonstrate attacks that can extract the secret weights of a BNN in a highly-parallelized hardware implementation.
    \item We formulate and implement the \emph{first} power-based side-channel countermeasures for neural networks by adapting masking to the case of neural networks. This process reveals new challenges and solutions to mask unique neural network computations that do not occur in the cryptographic domain.
    \item We validate both the insecurity of the baseline design and the security of the masked design using power measurement of an actual FPGA hardware and quantify the overheads of the proposed countermeasure.
\end{itemize}

We note that while there is prior work on theoretical attacks~\cite{ML-extraction-theory1,ML-extraction-theory2,ML-extraction-theory3,ML-extraction-theory4,ML-extraction-theory5} and digital side-channels~\cite{hua18,ML-extraction-SCA2,ML-extraction-SCA3,ML-extraction-SCA4} of neural networks, their physical side-channels are largely unexplored. Such research is needed because physical side-channels are orthogonal to these threats, fundamental to the Complementary Metal Oxide Semiconductor (CMOS) technology, and require extensive countermeasures as we have learned from the research on cryptographic implementations. A white paper is recently published on model extraction via physical side-channels~\cite{batina18}\footnote{After we submitted our work to HOST'20, that paper has been published at USENIX'19~\cite{batina19}.}.
This work does not study mitigation techniques and focuses on 8-bit/32-bit microcontrollers. We further analyze attacks on parallelized hardware accelerators and investigate the \emph{first} countermeasures.
\vspace{-0.15em}
\section{Threat Model and Relation to Prior Work}
\vspace{-0.15em}
\label{sec:threat}
This work follows the typical DPA threat model~\cite{Kocher2011}. The adversary has physical access to the target device or has a remote monitor~\cite{schellenberg18,zhao18,ramesh18} and obtains power measurements while the device processes secret information.
We assume that the security of the system is not based on the secrecy of the software or hardware design. 
This includes the details of neural network algorithm and its hardware implementation such as the data flow, parallelization and pipelining---in practice, those details are typically public but \textbf{what remains a secret is the model parameters obtained via training.}
For unknown implementations, adversary can use prior techniques to locate certain operations, which work in the context of physical~\cite{balasch15,eisenbarth10} and digital side-channels~\cite{zhang12,inci16} covering both hardware and software realizations.
This aspect of reverse engineering logic functions from bitstream~\cite{note08,benz12} is independent of our case and is generic to any given system.

\begin{figure}[t!]
	\begin{center}
		%\vspace{-2em}
		\hspace*{-.55em}
		\includegraphics[trim={0cm 0cm 0cm 0cm},clip,width=0.5\textwidth]{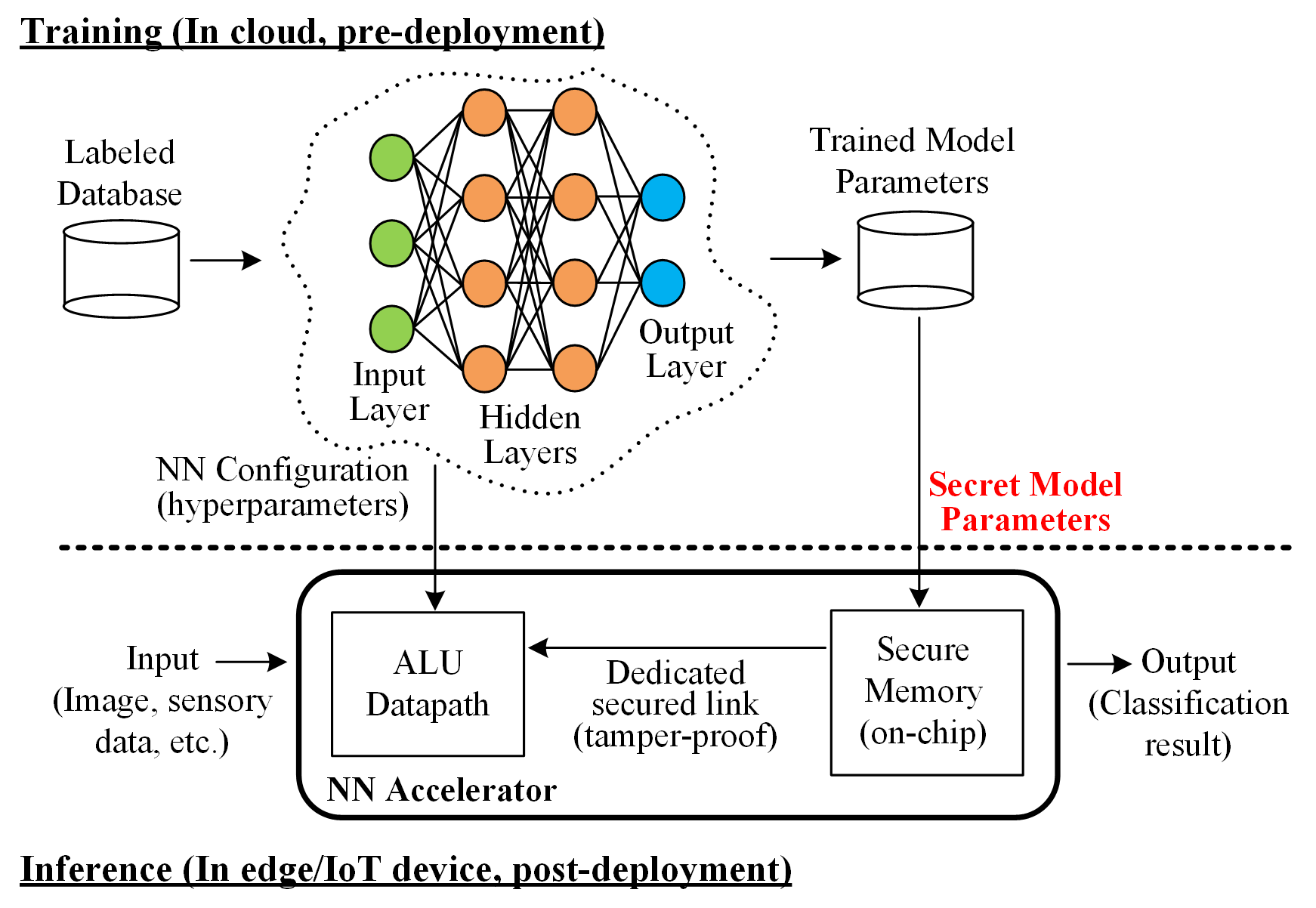}
	\end{center}
	\vspace{-1.6em}
	\caption{The adversary's goal is to extract the secret model parameters on the IoT/edge device during inference using side-channels.}
	\label{fig:threat-model}
	\vspace{-1em}
\end{figure}

Fig. \ref{fig:threat-model} outlines the system we consider.
The adversary in our model \textbf{targets the ML inference with the objective of learning the secret model parameters.} This is different than attacks on training set~\cite{shokri17} or the data privacy problem during inference~\cite{wei18}.
We assume the training phase is trusted but the obtained model is then deployed to operate in an untrusted environment.
Our attack is similar to a known-plaintext (input) attack and does not require knowing the inference output or confidence score, making it more powerful than theoretical ML extraction attacks~\cite{ML-extraction-theory1,ML-extraction-theory2,ML-extraction-theory3,ML-extraction-theory4,ML-extraction-theory5}. 

Since edge/IoT devices are the primary target of DPA attacks (due to easy physical access), we focus on BNNs that are suitable on such constrained devices~\cite{Courbariaux}.
A BNN also allows realizing the entire neural network on the FPGA without having external memory access.
Therefore, memory access pattern side-channel attacks on the network~\cite{hua18} cannot be mounted.
We furthermore consider digitally-hardened accelerator designs that execute in constant time and constant flow with no shared resources, disabling timing-based or other control-flow identification attacks~\cite{ML-extraction-SCA2,ML-extraction-SCA3,ML-extraction-SCA4}. 
This makes the attacks we consider more potent than prior work.
\vspace{-0.25em}
\section{BNN and the Target Implementation}
\vspace{-0.25em}
The following subsections give a brief introduction to BNN and discuss the details of the target hardware implementation.

\vspace{-0.5em}
\subsection{Neural Network Classifiers}
\vspace{-0.25em}
Neural networks consist of layers of neurons that take in an input vector and ultimately make a decision, e.g., for a classification problem.
Neurons at each layer may have a different use, implementing linear or non-linear functions to make decisions, applying filters to process the data, or selecting specific patterns.
Inspired from human nervous system, the neurons in a neural network transmit information from one layer to the other typically in a feed-forward manner.

\begin{figure}[t!]
	\begin{center}
		%\vspace{-2.3em}
		\includegraphics[trim={0cm 0cm 0cm 0cm},clip,width=0.48\textwidth]{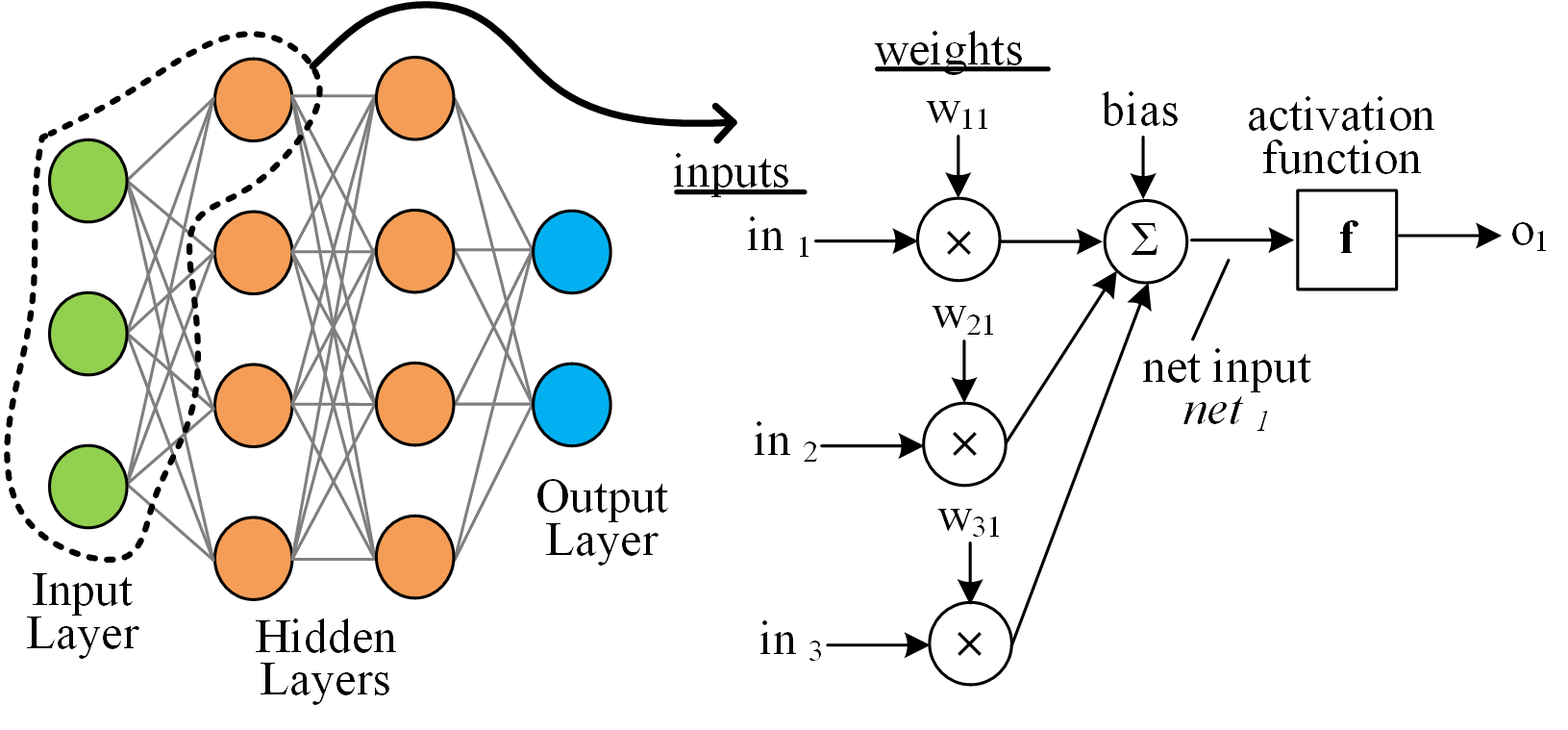}
	\end{center}
	\vspace{-1.5em}
	\caption{A simple neural network and a single neuron's function.}
	\label{fig:nn-basics}
	\vspace{-1.25em}
\end{figure}

Fig. \ref{fig:nn-basics} shows a simple network with two fully-connected hidden layers and depicts the function of a neuron. 
In a feed-forward neural network, each neuron takes the results from its previous layer connections, computes a weighted summation (row-reduction), adds a certain bias, and finally applies a non-linear transformation to compute its \emph{activation} output. The resulting activation value is used by the next layer's connected neurons in sequence. 
The connections between neurons can be strong, weak, or non-existent---the strength of these connections is called \emph{weight}, which is a critical parameter for the network.
The entire neural network model can be represented with these parameters, and with hyperparameters that are high-level parameters such as the number or type of the layers.

Neural networks have two phases: \emph{training} and \emph{inference}.
During training, the network self-tunes its parameters for the specific classification problem at hand.
This is achieved by feeding pre-classified inputs to the network together with their classification results, and by allowing the network to converge into acceptable parameters (based on some conditions) that can compute correct output values.
During inference, the network uses those parameters to classify new (unlabeled) inputs.

\vspace{-0.25em}
\subsection{BNNs}

A BNN works with binary weights and activation values. 
This is our starting point as the implementations of such networks have similarities with the implementation of block ciphers.
BNN reduces the memory size and converts a floating point multiplication to a single-bit XNOR operation in the inference~\cite{XNORNet}.
Therefore, such networks are suitable for constrained IoT nodes where some of the detection accuracy can be traded for efficiency.
Several variants of this low-cost approach exist to build neural networks with reasonably high accuracy~\cite{XNORNet,FINN,Courbariaux}.

\begin{figure}[t!]
\vspace{-1em}
\centering
  \includegraphics[scale=0.5]{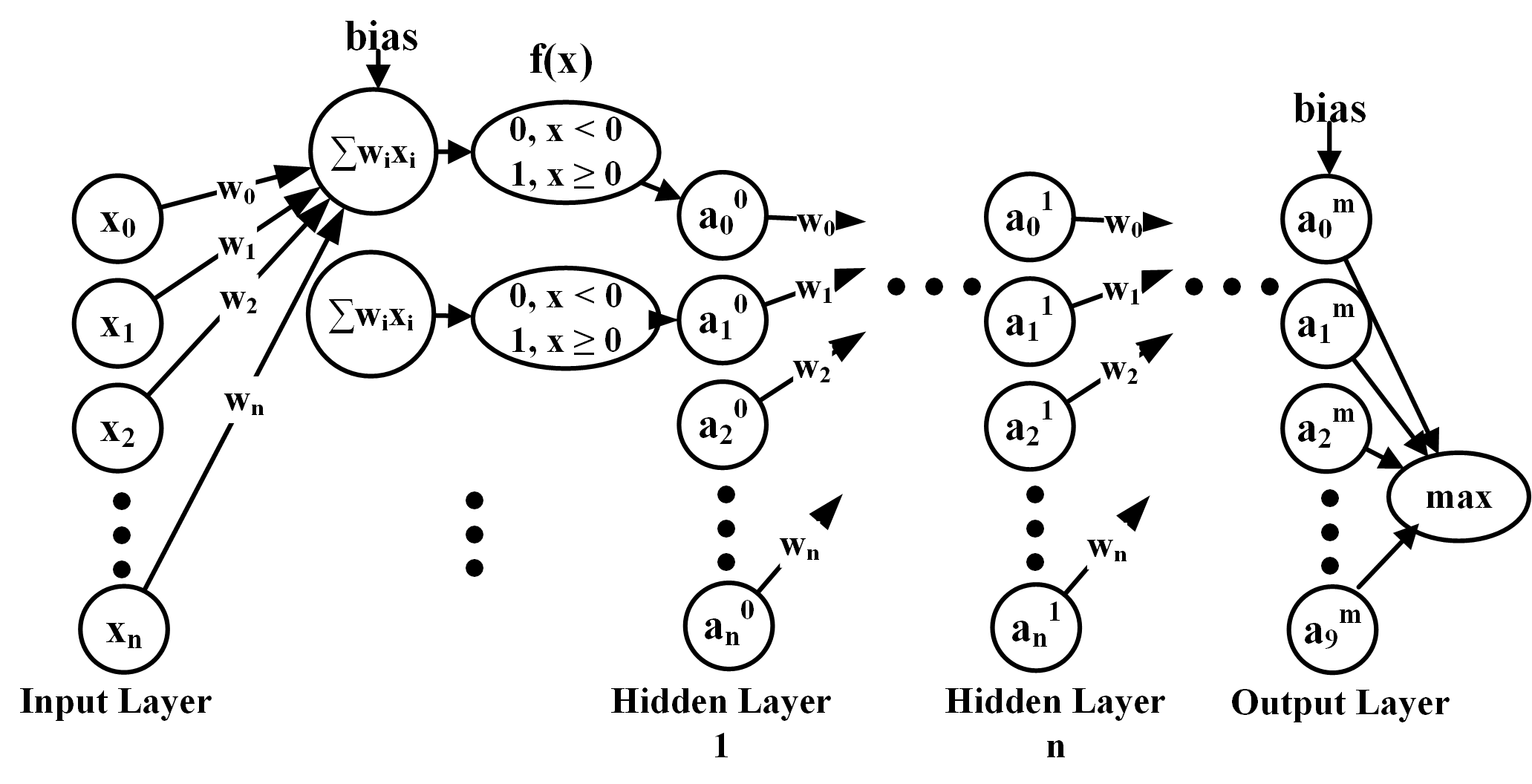}
  \vspace{-1em}
  \caption{Overview of the unprotected neural network inference. The adder tree first computes the weighted sum of input pixels. The activation function then binarizes the sum used by the next layer. Finally, the output layer returns the classification result by computing a maximum of last layer activations.}
  \vspace{-1em}
  \label{fig:overview}
\end{figure}

The following is the mathematical equation (1) for a typical neuron:
\vspace{-1em}

\begin{equation}
%\vspace{-1em}
\label{act}
    a = f(\sum w_ix_i + b)
\end{equation}

\noindent where $a$ is the activation value, $w_i$ is the weight, $x_i$ is the activation of the previous layer and $b$ is the bias value for the node. $w_i$, $x_i$, and $a$ have binary values of 0 and 1 respectively representing the actual values of -1 and +1. The function $f(x)$ is the non-linear activation function (2), which in our case is defined as follows:

\begin{equation}
\label{sign_fn}
    f(x) = \left\{\begin{array}{lr}
        0, & \text{for } x\leq 0\\
        1, & \text{for } x> 1\\
        \end{array}\right\}
 \end{equation} 
Equations (\ref{act}) and (\ref{sign_fn}) show that the computation involves a summation of weighted products with binary weights with a bias offset and an eventual binarization.

We build the BNN inference hardware which performs Modified National Institute of Standards and Technology (MNIST) classification (hand written digit recognition from 28-by-28 pixel images) using 3 feed-forward, fully-connected hidden layers of 1024 neurons. The implementation computes up to 1024 additions (i.e., the entire activation value of a neuron) in parallel.

\subsection{Unprotected Hardware Design}
\label{hw_design}

Fig. \ref{fig:overview} illustrates the sequence of operations in the neural network. One fully-connected layer has two main steps: (1) calculating the weighted sums using an adder tree and (2) applying the non-linear activation function $f$. To classify the image, the hardware sends out the node number with maximum sum in the output layer.

\begin{figure}[t!]
\vspace{-1em}
\centering
  \includegraphics[width=.5\textwidth]{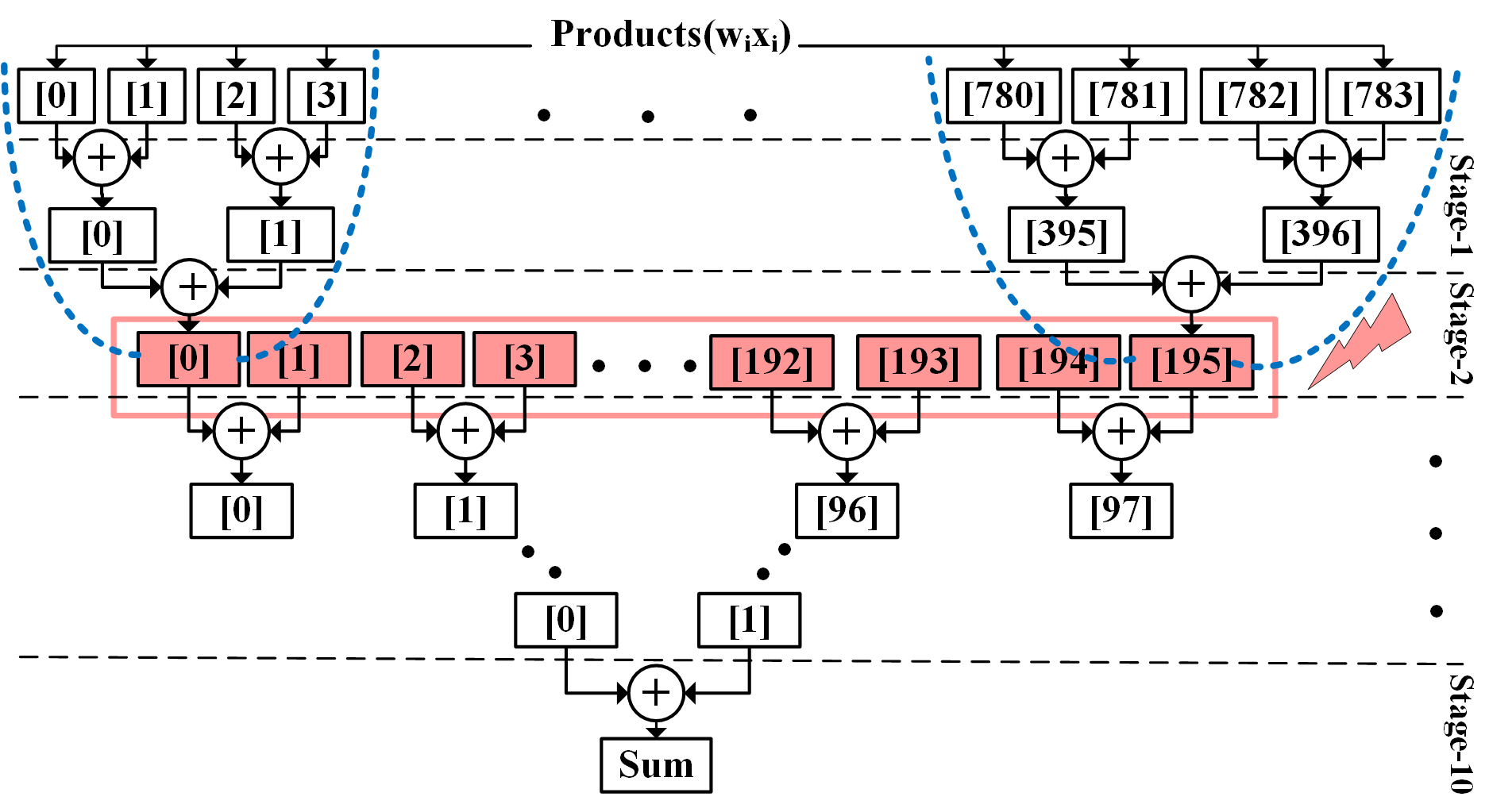}
  \vspace{-2em}
  \caption{Adder Tree used in HW Implementation. The figure shows the scenario where the 2nd stage registers(red) are targeted for DPA. This results in 16 possible key guesses corresponding to the 4 input pixels involved in the computation of each second stage register, grouped by the dotted blue line.}
  \label{fig:Addertree}
  \vspace{-1em}
\end{figure}

The input image pixels are first buffered or negated based on whether the weight is 1 or 0 respectively, and then fed to the adder tree shown in Fig. \ref{fig:Addertree}. We implemented a fully-pipelined adder tree of depth 10 as the hardware needs up to 1024 parallel additions to compute the sum. The activation function binarizes the final sum. After storing all the first layer activations, the hardware computes the second layer activations using the first layer layer activations. 
The output layer consists of 10 nodes, each representing a classification class (0-9).
The node index with the maximum confidence score becomes the output of the neural network.

The hardware reuses the same adder tree for each layer's computation, similar to a prior architecture~\cite{nurvitadhi16}.
Hence, the hardware has a throughput of approximately 3000 cycles per image classification. The reuse is not directly feasible as the adder tree (Fig. \ref{fig:Addertree}) can only support 784 inputs (of 8-bits) but it receives 1024 outputs from each of the hidden layers. Therefore, the hardware converts the 1024 1-bit outputs from each hidden layer to 512 2-bit outputs using LUTs. These LUTs take the weights and activation values as input, and produces the corresponding 512 2-bit sums, which is within the limits of the adder tree. Adding bias and applying the batch normalization is integrated to the adder tree computations. We adopt the batch-normalization free approach~\cite{BNFree}, the bias values are thus integer unlike the binary weights.

\section{An Example of DPA on BNN Hardware}
\label{sec:attack_and_results}
This section describes the attack we performed on the BNN hardware implementation that is able to extract secret weights. 

To carry out a power based side-channel attack, the adversary has to primarily focus on the switching activity of the registers, as they have a significant power consumption compared to combinational logic (especially in FPGAs).
The pipeline registers of the adder tree store the intermediate summations of the product of weights and input pixels. 
Therefore, the value in these registers is directly correlated to the secret---model weights in our case. 

Fig. \ref{fig:Addertree} shows an example attack. 4 possible values can be loaded in the output register $[0]$ of stage-1: $-[0]-[1]$, $-[0]+[1]$, $[0]-[1]$ and $[0]+[1]$ corresponding to the weights of (0,0), (0,1), (1,0) and (1,1), respectively.
These values will directly affect the computation and the corresponding result stored in stage-2 registers. Therefore, a DPA attack with known inputs ($x_i$) on stage-2 registers (storing $w_ix_i$ accumulations) can reveal 4-bits of the secret weights ($w_i$).
The attack can target any stage of the adder tree but the number of possible weight combinations grows exponentially with depth.

\begin{figure}[t!]
\vspace{-1em}
\centering
  \hspace*{-2em}
  \includegraphics[width=0.45\textwidth]{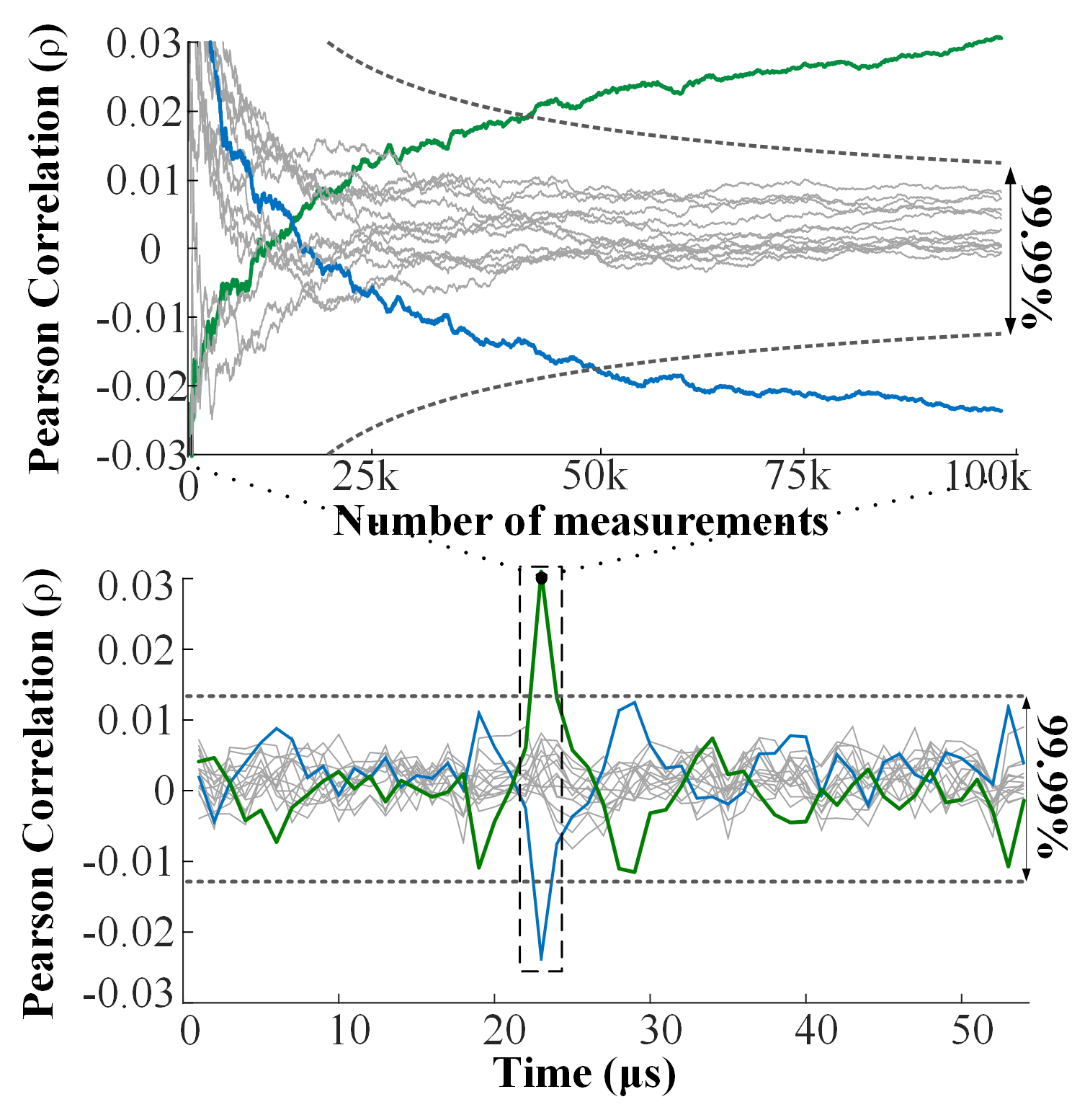}
  \vspace{-1.2em}
  \caption{Pearson Correlation Coefficient versus time and number of traces for DPA on weights. Lower plot shows a high correlation peak at the time of target computation, for the correct weight guess denoted in green. The upper plot shows that approximately 40k traces are needed to get a correlation of 99.99\% for the correct guess. The confidence intervals are shown in dotted lines. The blue plot denotes the 2's complement of the correct weight guess}
  \label{fig:DPA_adder}
  \vspace{-1.5em}
\end{figure}

Since the adder tree is pipelined, we need to create a model based on hamming distance of previous cycle and current cycle summations in each register. 
To aid the attack, we developed a cycle-accurate hamming-distance simulator for the adder pipeline. It first computes the value in each register every cycle for all possible weight combinations given a fixed input. Next, it calculates the hamming distance of individual registers for each cycle using the previous and current cycle values. Finally, it adds the hamming distances of all the registers for a cycle to model the overall power dissipation for that cycle.

Fig. \ref{fig:DPA_adder} illustrates the result of the attack on stage-2 registers.
There is a strong correlation between the correct key guess and the power measurements crossing the 99.99\% confidence threshold after 45k measurements.
The false positive leak is due to signed multiplication and is caused by the additive inverse of the correct key, which is expected and thus does not affect the attack.
Using this approach, the attacker can successively extract the value of weights and biases for all the nodes in all the layers, starting from the first node and layer. 
The bias, in our design, is added after computing the final sum in the 10$^{th}$ stage, before sending the result to the activation function. Therefore the adversary can attack this addition operation by creating a hypothesis for the bias. Another way extract bias is by attacking the activation function output since the sign of the output correlates to the bias.
\section{Side-Channel Countermeasures}
\label{sec:defense}
This section presents our novel countermeasure against side-channel attacks.
The development of the countermeasure highlights unique challenges that arise for masking neural networks and describes the implementation of the entire neural network inference.

Masking works by making \emph{all} intermediate computations independent of the secret key---i.e., rather than preventing the leak, by encumbering adversary's capability to correlate it with the secret value.
The advantage of masking is being implementation agnostic.
Thus, they can be applied on any given circuit style (FPGA or ASIC) without manipulating back-end tools but they require algorithmic tuning, especially to mask unique non-linear computations.

The main idea of masking is to split inputs of all key-dependent computations into two randomized shares: a one-time random mask and a one-time randomized masked value. 
These shares are then independently processed and are reconstituted at the final step when the final output is generated. 
This would effectively thwart first-order side-channel attacks probing a single intermediate computation. 
Higher-order attacks probing multiple computations~\cite{meserges2000}---masks and masked computations---can be further mitigated by splitting inputs of key-dependent operations into more shares~\cite{akkar2003}. 
Our implementation is designed to be first-order secure but can likewise be extended for higher order attacks.

Fig. \ref{fig:overview} highlights that a typical neural network inference can be split into 3 distinct types of operations: (1) the adder tree computations, (2) the activation function and the (3) output layer max function. Our goal is to mask these functions to increase resilience against power side-channel attacks. These specific functions are unique to a neural network inference. 
Hence, we aim to construct novel masked architectures for them using the lessons learned from cryptographic side-channel research.
We will explain our approach in a \emph{bottom-up fashion} by describing masking of individual components first, and the entire hardware architecture next.

\subsection{Masking the Adder Tree}
\label{adder_tree_masking}
Using the approach in Fig. \ref{fig:Addertree}, the adversary can attack any stage of the adder tree to extract the secret weights. 
Therefore, the countermeasure has to break the correlation between the summations generated at each stage and the secret weights. We use the technique of message blinding to mask the input of the adder tree. 

\begin{figure}[t!]
\centering
  \includegraphics[scale=.8]{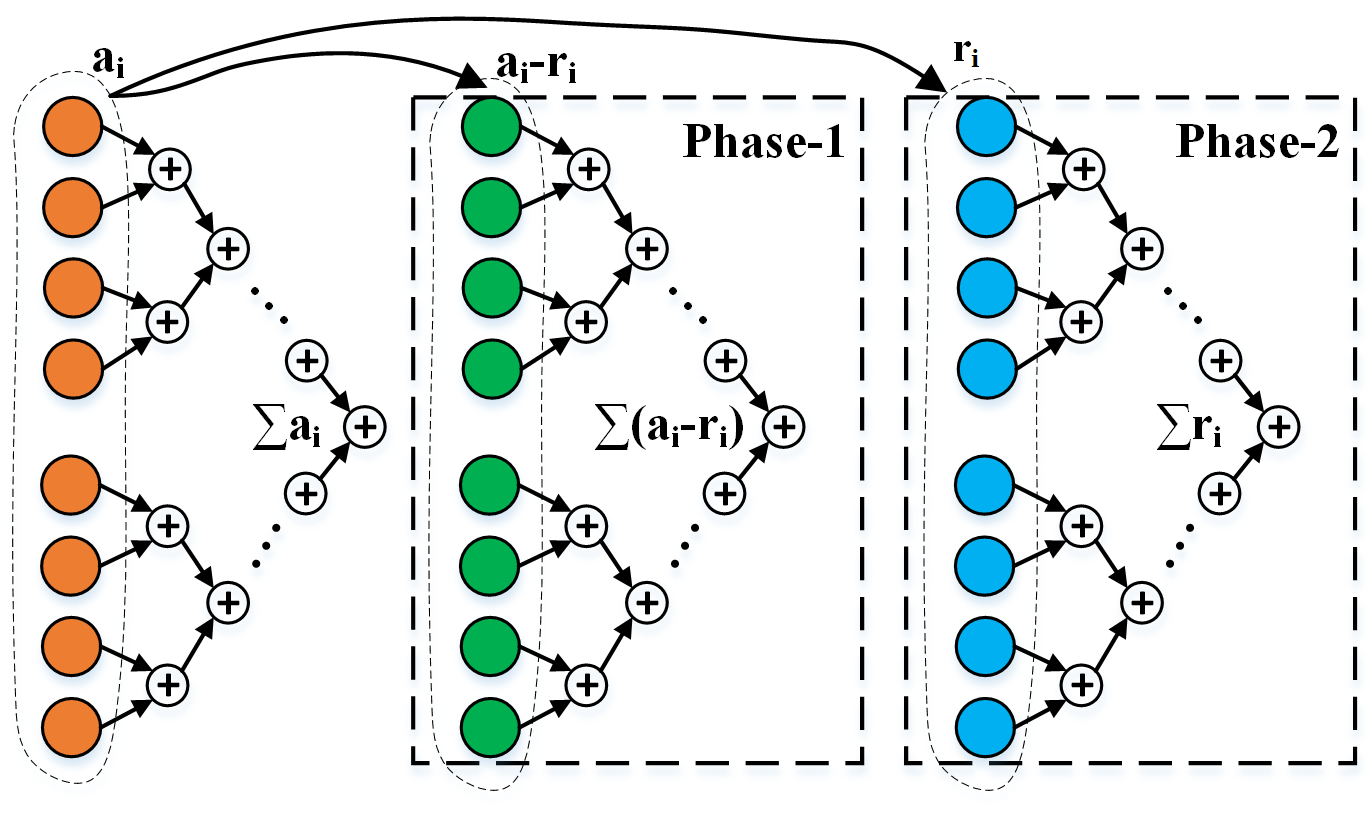}
  \vspace{-2.5em}
  \caption{Masking of adder tree. Each input pixel depicted in orange is split into two arithmetic shares depicted by the green and blue nodes with unique random numbers ($r_i$s). The masked adder tree computes branches sequentially.}
  \label{fig:masked_adder}
  \vspace{-1em}
\end{figure}

Blinding is a technique where the inputs are randomized before sending to the target unit for computation. This prevents the adversary from knowing the actual inputs being processed, which is usually the basis for known-plaintext power-based side channel attacks.
Fig. \ref{fig:masked_adder} shows our approach, that uses this concept by splitting each of the 784 input pixels $a_i$ into two arithmetic shares $r_i$ and $a_i\!-\!r_i$, where each $r_i$ is a unique 8-bit random number. These two shares are therefore independent of the input pixel value, as $r_i$ is a fresh random number never reused again for the same node. The adder tree can operate on each share individually due to additive homomorphism---it generates the two final summations for each branch such that their combination (i.e., addition) will give the original sum. 
Since the adder tree is reused for all layers, hardware simply repeats the masking process for subsequent hidden layers using fresh randomness in each layer.  

\subsubsection{A Unique and Fundamental Challenge for Arithmetic Masking of Neural Networks}
\label{msbprob}
The arithmetic masking extension to the adder tree is unfortunately non-trivial due to differences in the fundamental assumptions. 
Arithmetic masking aims at decorrelating a variable $x$ by splitting it into two statistically independent shares: $r$ and $(x-r)$ $mod$ $k$. The modulo operation in $(x-r)$ $mod$ $k$ exists in most cryptographic implementations because most of them are based on finite fields. In a neural network, however, subtraction means computing the actual difference \emph{without} any modulus. This introduces the notion of sign in numbers, which is absent in modulo arithmetic, and is the source of the problem.

Consider two uniformly distributed 8-bit unsigned numbers $a$ and $r$. In a modulo subtraction, the result will be $(a-r)\;mod\;256$, which is again an 8-bit unsigned number lying between 0 and 255. In an actual subtraction, however, the result will be $(a-r)$, which is a 9-bit number with MSB being the sign bit. 

\begin{table}[h!]
\vspace{-1.25em}
\centering
\caption{Probability of $a-r$ being positive or negative}
\vspace{-0.5em}
\begin{tabular}{ |c|c|c| } 
 \hline
 Scenario & Positive & Negative \\ 
 \hline
 \hline
 $a>128\,\&\,r>128$ & 50\% & 50\% \\ 
 \hline
 $a>128\,\&\,r<128$ & 100\% & 0\% \\ 
 \hline
 \hline
 $a<128\,\&\,r>128$ & 0\% & 100\% \\ 
 \hline
 $a<128\,\&\,r<128$ & 50\% & 50\% \\ 
 \hline
\end{tabular}
\label{tab_prob}
\vspace{-0.75em}
\end{table}

Table \ref{tab_prob} lists four possible scenarios of arithmetic masking based on the magnitude of the two unsigned 8-bit shares. In a perfect masking scheme, probability of $a-r$ being either positive or negative should be 50\%, irrespective of the magnitude of the input $a$. Let's consider the case when $a>128$, which has a probability of 50\%. If $r<128$, which also has a 50\% probability, the resulting sum $a-r$ is always positive. Else if $r>128$, the value $a-r$ can both be positive or negative with equal probabilities due to uniform distribution. 
Therefore, given $a>128$, the probability of the arithmetic mask being positive is $(50+25)\%=75\%$ and being negative is $25\%$.
Table \mbox{\ref{tab_prob}} lists the other case when $a<128$, which results a similar correlation between $a$ and $a-r$.
This is showing a clear information leak through the sign bit of arithmetic masks.

The discussed vulnerability does not happen in modulo arithmetic as \emph{there is no sign bit; the modulo operation wraps around the result if it is out of bounds, to obey the closure property}. Evaluating the correlation of $(a+r)$ instead of $(a-r)$ yields similar results. Likewise, shifting the range of $r$ based on $a$, to uniformly distribute $(a-r)$ between -128 to 127, would not resolve the problem and further introduces a bias in both shares. 

\subsubsection{Addressing the Vulnerability with Hiding}
The arithmetic masking scheme can be augmented to decorrelate the sign bit from the input. 
We used hiding to address this problem.
We used hiding \emph{just for the sign bit computation}. 
Hiding techniques target constant power consumption, irrespective of the inputs, which makes it harder for an attacker to correlate the intermediate variables. 
Power equalized building blocks using techniques like Wave Differential Dynamic Logic (WDDL)~\mbox{\cite{WDDL}} can achieve close to a constant power consumption to mitigate the vulnerability.

The differential part of WDDL circuits aims to make the power consumption constant throughout the operation, by generating the complementary outputs of each gate along with the original outputs. Differential logic makes it difficult for an attacker to distinguish between a $0 \rightarrow 1$ and a $1 \rightarrow 0$ transition, however, an attacker can still distinguish between a $0 \rightarrow 0$ and a $0 \rightarrow 1$ transition or a $1 \rightarrow 1$ and a $1 \rightarrow 0$ transition. Therefore, the differential logic alone is still susceptible to side channel leakages, as the power activity is easily correlated to the input switching pattern. This vulnerability is reduced by dynamic logic, where all the gates are pre-charged to 0, before the actual computation.

We use the WDDL gates to solve our problem of sign bit leakages, by modifying the adders to compute the sign bit in WDDL style.
Following is the equation of the addition, when two 8-bit signed numbers $a$ and $b$, represented as $(a_7 a_6\cdot\cdot\cdot a_0)$ and $(b_7 b_6\cdot\cdot\cdot b_0)$ are added to give a 9-bit signed sum $s$ represented by $(s_8 s_7\cdot\cdot\cdot s_0)$:
\begin{equation}
\label{eq:wddl1}
s = a + b
\end{equation}
After sign-extending $a$ and $b$,
\begin{equation}
\label{eq:wddl2}
\{s_8 s_7 s_6 \cdot\cdot\cdot s_0\} = \{a_7 a_7 a_6\cdot\cdot\cdot a_0\}+\{b_7 b_7 b_6\cdot\cdot\cdot b_0\}
\end{equation}
Performing regular addition on the leftmost 8 bits of $a$ and $b$, and generating a carry $c$, the equation of $s_8$ becomes
\begin{equation}
\label{eq:wddl3}
s_8 = a_7 \oplus b_7 \oplus c
\end{equation}
Expanding the above expression in terms of AND, OR and NOT operators results:
\begin{equation}
\label{eq:wddl4}
s_8 = (\overline{a_7}\cdot b_7 \cdot \overline{c}) | (a_7 \cdot\overline{b_7}\cdot\overline{c}) | (a_7\cdot b_7\cdot c) | (\overline{a_7}\cdot\overline{b_7}\cdot c)
\end{equation}
Representing the expression only in terms of NAND, so that we can replace all the NANDs by WDDL NAND gates reveals:
\begin{equation}
\label{eq:wddl5}
s_8 = \overline{\overline{(\overline{a_7}\cdot b_7\cdot\overline{c})}\cdot \overline{(a_7\cdot\overline{b_7}\cdot\overline{c}}\cdot\overline{ (a_7\cdot b_7\cdot c)}\cdot\overline{ (\overline{a_7}\cdot\overline{b_7}\cdot c)}}
\end{equation}

\begin{figure}[t!]
\centering
  \includegraphics[width=.5\textwidth]{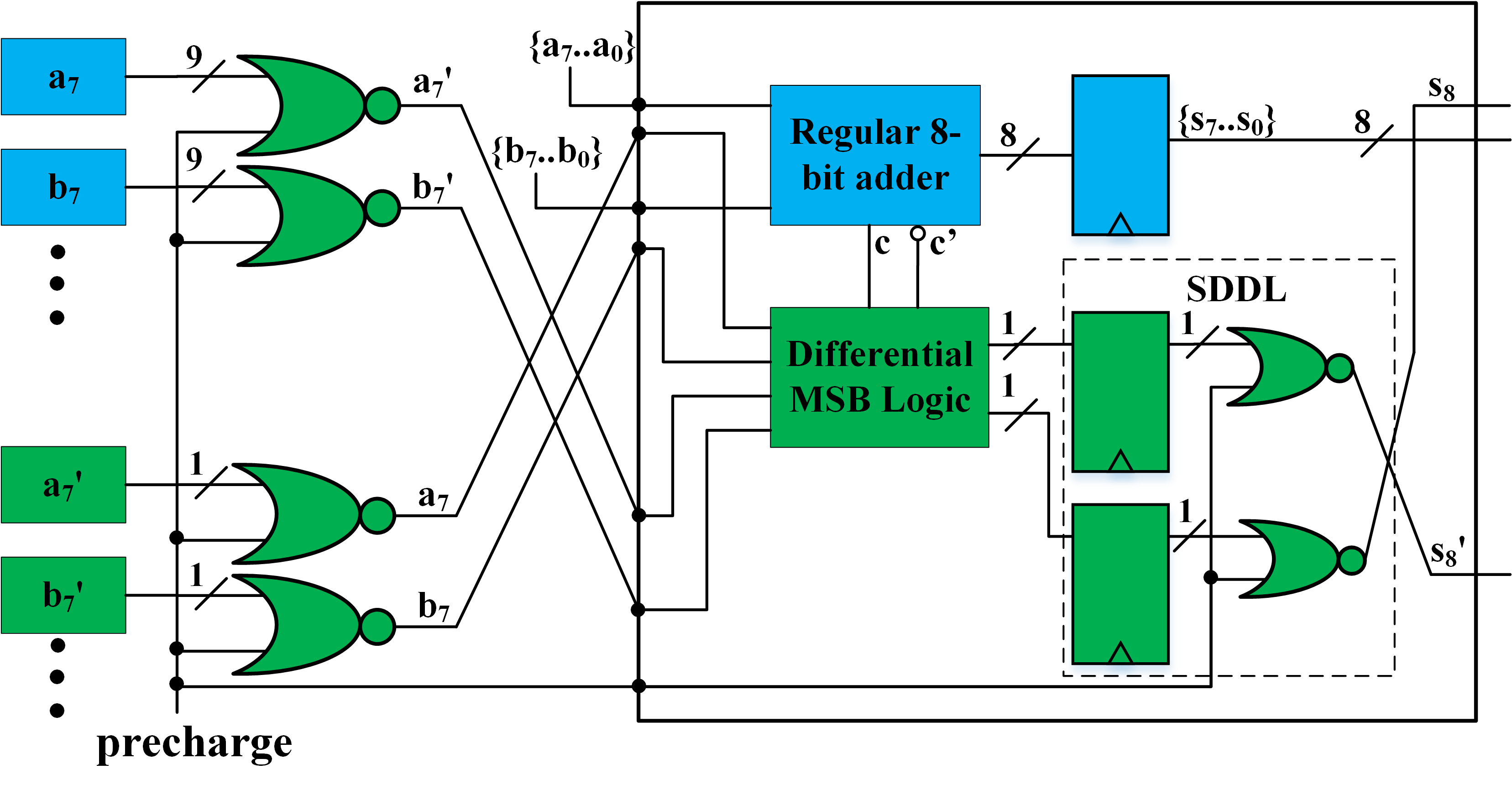}
  \vspace{-2em}
  \caption{Circuit diagram of the proposed adder with MSB computed in WDDL style as described in Eq.(\ref{eq:wddl1})-(\ref{eq:wddl5}). Each of the 784 arithmetic shares ($a,b,...$) are fed to these adders. All the bits except the MSB undergo a regular addition. The MSBs of the two operands along with the generated carry are fed to the Differential MSB Logic block, which computes the MSB and its complement by replacing the NAND gates in Eq (\ref{eq:wddl5}) by WDDL gates. The pipeline registers in the tree are replaced by SDDL registers. The NOR gates generate the pre-charge wave at the start of the logic cones. The regular and the WDDL specific blocks are depicted in blue and green respectively}
  \label{fig:wddl_adder}
  \vspace{-1.5em}
\end{figure}

Fig. \ref{fig:wddl_adder} depicts the circuit diagram for the above implementation. The WDDL technique is applied to the MSB computation by replacing each NAND function in Eq (\ref{eq:wddl5}) with WDDL NAND gates. 
The pipeline registers of the adder tree are replaced by Simple Dynamic Differential Logic (SDDL) registers~\cite{WDDL}. Each WDDL adder outputs the actual sum $s$ and the complement of its MSB $s_8'$, which go as input to the WDDL adder in the next stage of the pipelined tree. Therefore, we construct a resilient adder tree mitigating the leakage in the sign-bit.

\subsection{Masking the Activation Function}
\label{act_func}

The binary sign function (Eq. \ref{sign_fn}) is the activation function of BNN. This function generates +1 if the weighted sum is positive, else -1 if the sum is negative. In the unmasked implementation, the sign function receives the weighted sum of the 784 original input pixels, whereas in the masked implementation, it receives the two weighted sums corresponding to each masked share. So, the masked function has to compute the sign of the sum of two shares \emph{without} actually adding them. Using the fact that the sign only depends on the MSB of the final sum, we propose a novel masked sign function that sequentially computes and propagates the masked carry bits in a ripple carry fashion.

\begin{figure}[t!]
\centering
  \includegraphics[width=0.48\textwidth]{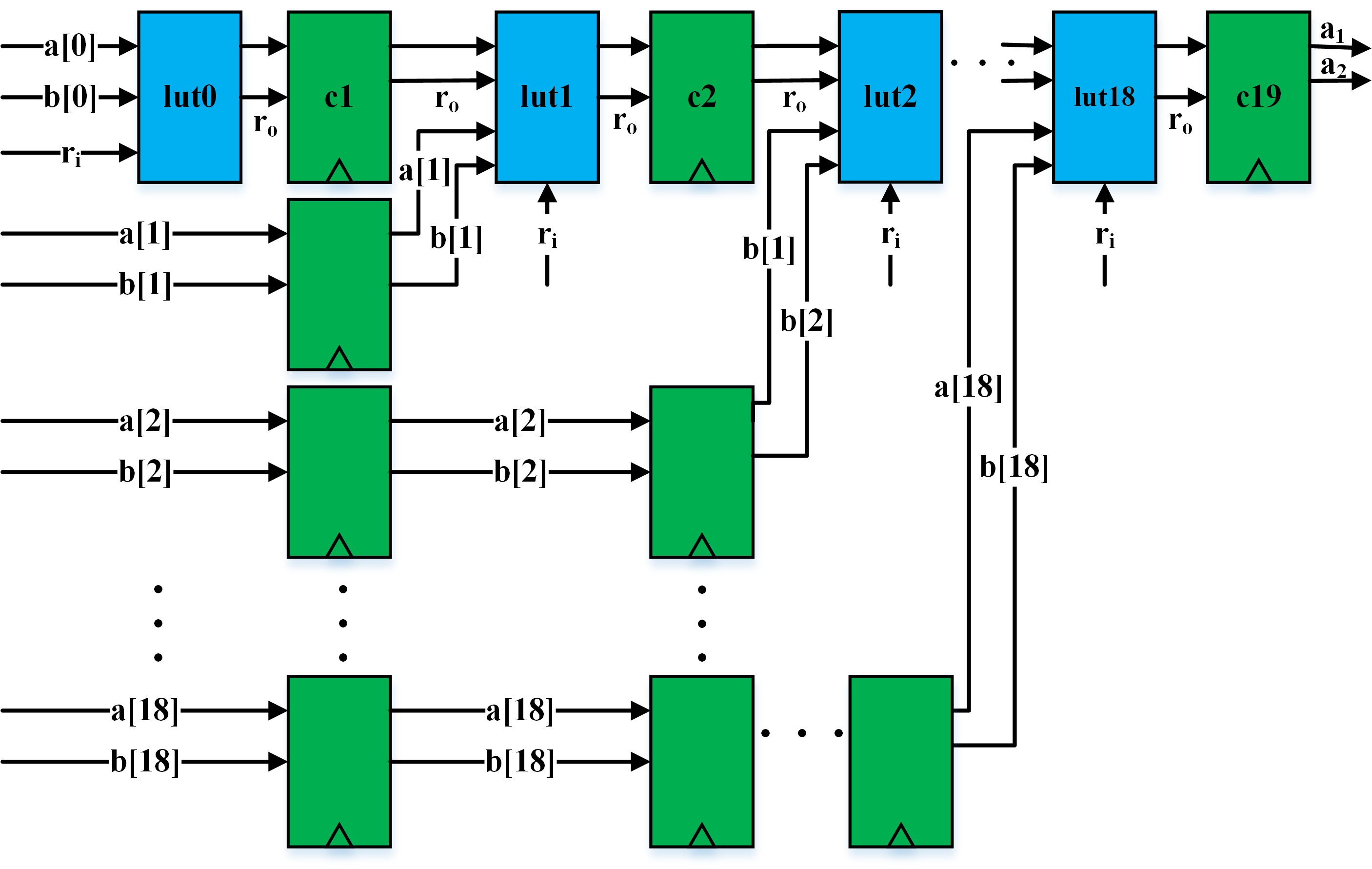}
  \vspace{-1em}
  \caption{Hardware design of the masked binarizer. It comprises of a chain of LUTs (lut0-lut18) denoted in blue, computing the carry in ripple carry fashion. Each LUT is masked by a fresh random number ($r_i$). The whole design is fully pipelined to maintain the original throughput by adding flip flops (in green) at each stage. 
  }
  \label{fig:masked_carrygen}
  \vspace{-1.5em}
\end{figure}

Fig. \ref{fig:masked_carrygen} shows the details of our proposed masked sign function hardware.
This circuit generates the first \emph{masked carry} using a Look-up-Table (LUT) that takes in the LSB of both shares and a  fresh random bit ($r_i$) to ensure the randomization of the intermediate state, similar in style to prior works on masked LUT designs \cite{reparaz15}.
LUT function computes the masked function with the random input and generates two outputs: one is the bypass of the random value ($r_i$) and the other is the masked output ($r_i\oplus f(x)$) where $f(x)$ is the carry output.
The entire LUT function for each output can fit into a single LUT to reduce the effects of glitches~\cite{nikova06}. To further reduce the impact of glitches, the hardware stores each LUT output in a flip-flop. These are validated empirically in Section \ref{sec:eval}.

The outputs of an LUT are sent to the next LUT in chain and the next masked carry is computed accordingly. 
From the second LUT and onward, each LUT has to also take the masked carry and mask value generated from the prior step. The output $r_o$ is simply the input $r_i$ like a forward bypass, because the mask value is also needed for the next computation.
This way the circuit processes all the bits of the shares and finally outputs the last carry bit which decides the sign of the sum. Each LUT computation is masked by a fresh random number.
More efficient designs may also be possible using a masked Kogge-Stone adder (e.g., by modifying the ones described in \cite{Schneider_arithmeticaddition}).

Fig. \ref{fig:masked_carrygen} illustrates that the first LUT is a 3-bit input 2-bit output LUT because there is no carry-in for LSB, and all the subsequent LUTs have 5-bit inputs and 2-bit outputs since they also need previous stage outputs as their inputs.
After the final carry is generated, which is also the sign bit of the sum, the hardware can binarize the sum to 1 or 0 based on whether the sign bit is 0 or 1 respectively. This is the functionality of the final LUT, which is different from the usual masked carry generators in the earlier stages.

The circuit has 19 LUTs in serial; each LUT output is registered for timing and side-channel resilience against glitches. This design, however, adds a latency of 19 cycles to compute each activation value, increasing the original latency. Therefore, instead of streaming each of the 19 bits on the top row of LUTs sequentially in Fig. \ref{fig:masked_carrygen}, the entire 19 bit sum is registered in the first stage, and each bit is used sequentially throughout the 19 cycles. This avoids the 19 cycle wait time for consecutive sums and brings back the throughput to 1 activation per cycle.

\begin{figure*}[t!]
  \vspace{-1em}
  \centering
  \includegraphics{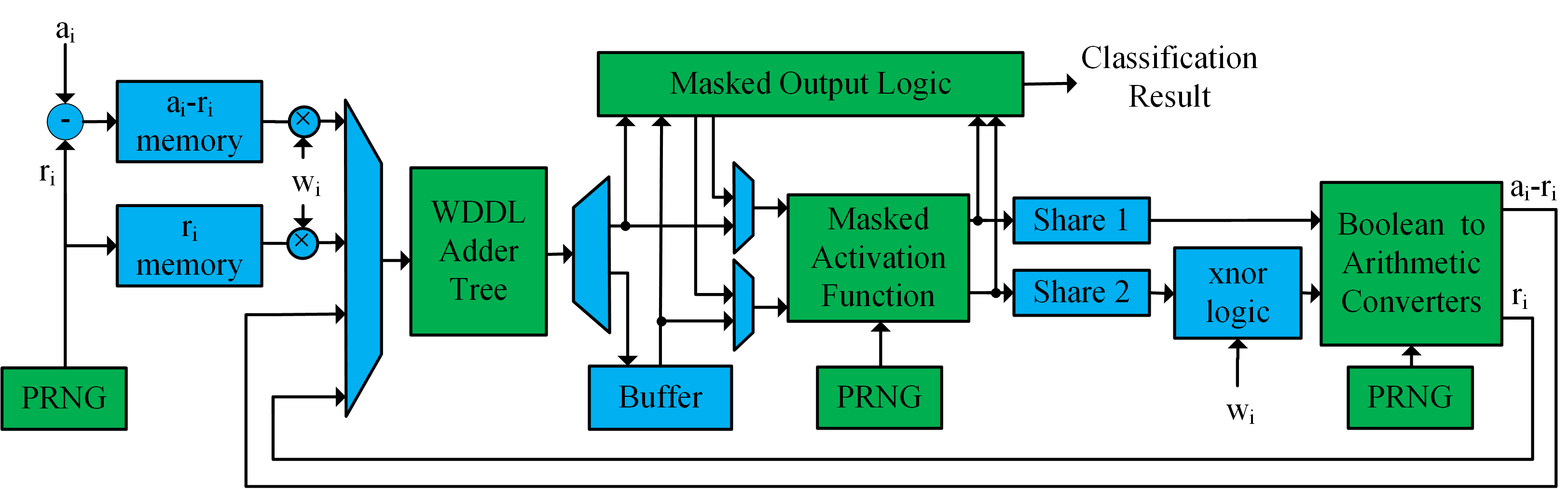}
  \vspace{-1em}
  \caption{Components of the masked BNN. Blocks in green represent the proposed masking blocks.}
  \label{fig:masked_bnn_full}
  \vspace{-2em}
\end{figure*}

\subsection{Boolean to Arithmetic Share Conversion}
Each layer generates 1024 pairs of Boolean shares, which requires two changes in the hardware. First, the adder tree supports 784 inputs which cannot directly process 1024 shares. Second, the activation values are in the form of two Boolean shares while the masking of adder tree requires arithmetic shares as discussed in Section \ref{adder_tree_masking}. Using the same strategy of the unmasked design, the hardware adds 1024 1-bit shares pairwise to produce 512 2-bit shares before sending them to the adder tree. To resolve the conversion of Boolean to arithmetic conversion, the hardware can generate $R$ such that
\begin{equation}
x = x_1 \oplus x_2
\end{equation}
\begin{equation}
x = R + x_2
\end{equation}

Using masked LUTs, the hardware performs signed addition of 1024 shares to 512 shares, and it also produces the arithmetic shares. The LUTs take in two consecutive activation values already multiplied by the corresponding weights, and a 2-bit signed random number to generate the arithmetic shares. Since multiplication in binary neural network translates to an XNOR operation~\cite{XNORNet}, the hardware performs an XNOR operation using the activation value with its corresponding weight before sending it to the LUT. Since the activation value is in the form of 2 Boolean shares, the hardware performs XNOR only on one of the shares as formulated below:
\begin{equation}
    a\,\overline{\oplus}\,w = b
\end{equation}
\begin{equation}
    a = a_1\,\oplus\,a_2
\end{equation}
\begin{equation}
   (a_1\,\oplus\,a_2)\,\overline{\oplus}\,w\,=\,(a_1\,\overline{\oplus}\,w)\,\oplus\,a_2 
\end{equation}

The LUTs have five inputs: two shares that are not XNORed, two shares that are XNORed and a 2-bit signed random number. If the actual sum of the two consecutive nodes is \emph{a\textsubscript{i}}, then the LUT outputs \emph{r\textsubscript{i}} and \emph{a\textsubscript{i}-r\textsubscript{i}} range from -2 to +1 since it is a 2-bit signed number and weighted sum of two nodes will range from -2 to +2. 
Therefore, \emph{a\textsubscript{i}-r\textsubscript{i}} can range from -3 to +4 and should be 4-bit wide. The hardware has 512 LUTs that convert the 1024 pairs of Boolean shares to 512 pairs of arithmetic shares. After the conversion, the hardware reuses the same adder tree masking that was described in Section \ref{adder_tree_masking}. The arithmetic shares have a leakage in MSB as discussed in Subsection \ref{msbprob}, but because the same WDDL style adder tree is reused, this is addressed for all layers.

\subsection{Output Layer}
\label{outfn}
In the output layer, for an unmasked design, the index of the node with maximum confidence score is generated as the output of the neural network. In the masked case, however, the confidence values are split in two arithmetic shares, which by definition cannot be combined. Equations (14--16) formulate the masking operations of the output layer. Basically, we check if the sum of two numbers is greater than the sum of another two numbers, without looking at the terms of each sum at the same time. Therefore, instead of adding the two shares of the confidence values and comparing them, we subtract one share of a confidence value from another share of the other confidence value. This still solves the inequality, but looks at the shares of two different confidence scores.
\begin{equation}
    a_1 + a_2 \geq b_1 + b_2 
\end{equation}
\begin{equation}
    a_1 - b_2 \geq b_1 - a_2
\end{equation}
\vspace{-1.75em}
\begin{equation}
\label{eq:outfn_eqn}
    (a_1 - b_2) + (a_2 - b_1) \geq 0
\end{equation}

This simplifies the original problem to the previous problem of finding the sign of the sum of two numbers without combining them. Hence, in the final layer computation, the hardware reuses the masked carry generator explained in Section \ref{act_func}. 

\subsection{The Entire Inference Engine}
Fig. \ref{fig:masked_bnn_full} illustrates all the components of the masked neural network. 
First, the network arithmetically masks the input pixel $a_i$ using fresh $r_i$ generated by the PRNG. Next, the WDDL style adder tree processes each of the masks ($r_i$) and the masked values ($a_i-r_i$) in two sequential phases. The demultiplexer at the adder tree output helps to buffer the first phase final summations, and pass the second phase summations to the masked activation function directly. The masked activation function produces the two Boolean shares of the actual activation using fresh randomness from the PRNG. The network XNORs one share with the weights and sends the second share directly to the Boolean to arithmetic converters. The converters produce 512 arithmetic shares from the 1024 Boolean shares using random numbers generated by the PRNG. The hardware feeds the arithmetic shares $a_i-r_i$ and $r_i$ to the adder tree and repeats the whole process for each layer.
Finally, the output layer reuses the masked activation function to find the node with maximum confidence from the arithmetic shares of the confidence values and computes the classification result output as described in Subsection \ref{outfn}.
\begin{figure*}[t!]
\centering
  \vspace{-.25em}
  \includegraphics[scale=0.89]{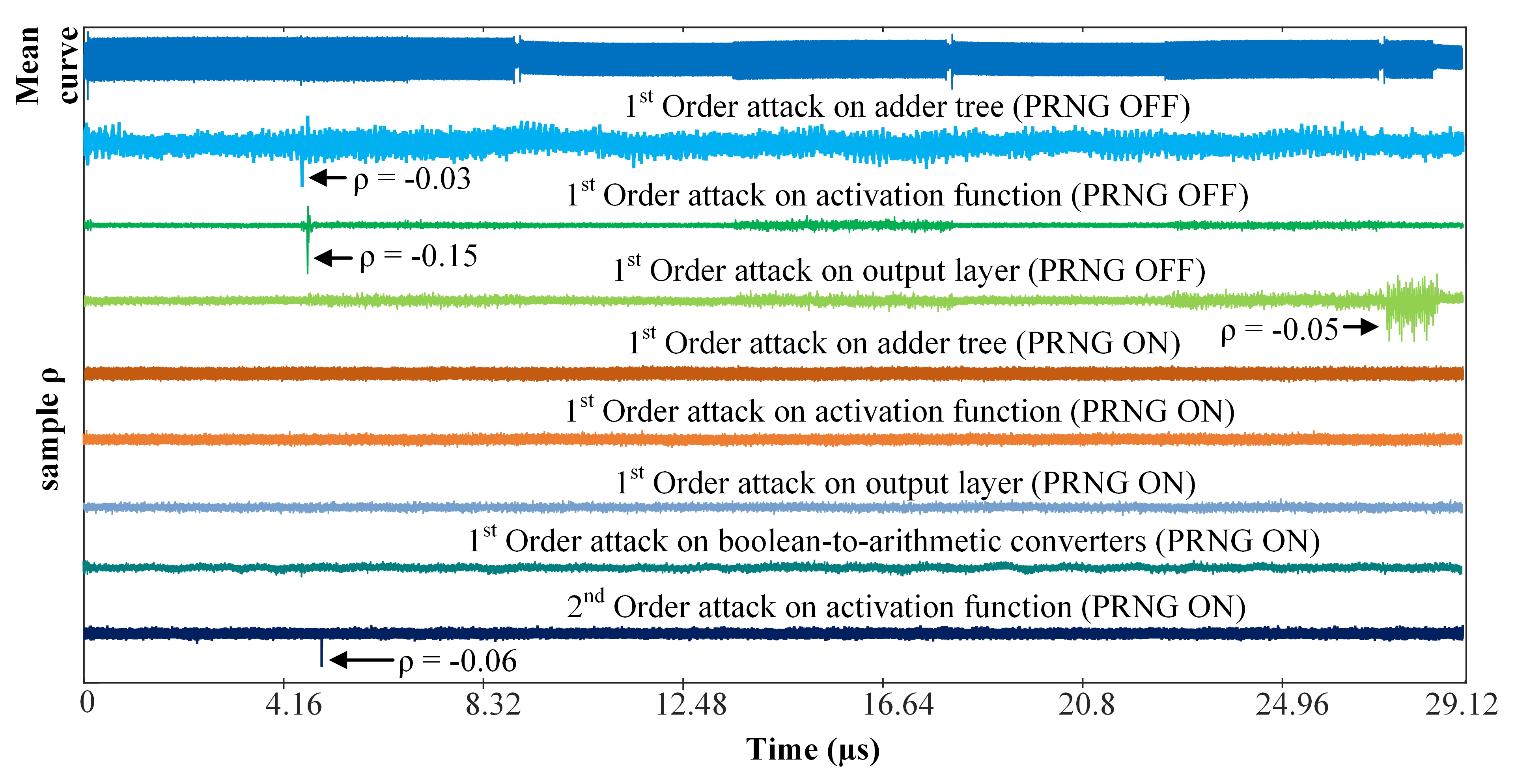}
  \vspace{-1.25em}
  \caption{Side-channel evaluation tests. First-order attacks on the unmasked design (PRNG off) show that it leaks information while the masked design (PRNG on) is secure. The second-order attack on the masked design succeeds and validates the number of measurements used in the first-order attacks.}
  \label{fig:res1}
  %\vspace{-1em}
\end{figure*}

\section{Leakage and Overhead Evaluation}
\label{sec:eval}
This section describes the measurement setup for our experiments, the evaluation methodology used to validate the security of the unprotected and protected implementations, and the corresponding results with overhead quantification.

\subsection{Hardware Setup}
Our evaluation setup used the SAKURA-X board \cite{sakurax}, which includes a Xilinx Kintex-7 (XC7K160T-1FBGC) FPGA for processing and enables measuring the voltage drop on a \mbox{10m\si{\ohm}} shunt-resistance while making use of the on-board amplifiers to measure FPGA power consumption.
The clock frequency of the design was 24MHz. We used the Picoscope 3206D oscilloscope to take measurements with the sampling frequency set to 250MHz. To amplify the output of the Kintex-7 FPGA, we used a low-noise amplifier provided by Riscure (HD24248) along with the current probe setup. The experiment can also be conducted at a lower sampling rate by increasing the number of measurements~\mbox{\cite{leakynoise}}.

\subsection{Side-Channel Test Methodology}
Our leakage evaluation methodology is built on the prior test efforts on cryptographic implementations \mbox{\cite{reparaz-mask-16,balasch15,reparaz15}}.
We performed DPA on the 4 main operations of an inference engine as stated before, viz. adder tree, activation function, Boolean to arithmetic share conversion, and output layer max function. 
Pseudo Random Number Generators (PRNG) produced the random numbers required for masking---any cryptographically-secure PRNG can be employed to this end.  
We first show the first-order DPA weight recovery attacks on the masked implementation with PRNG disabled. With PRNG off, the design's security is equivalent to that of an unmasked design. We illustrate that such a design leaked information for all the three operations, which ensured that our measurement setup and recovery code was sound. Next, we turned on the PRNG and performed the same attack which failed for all the operations. 
We also performed a second-order attack to validate the number of traces used in the first-order analysis.
The power model was based on hamming distance of registers that was generated using our HD simulator for the neural network and the tests used the Pearson correlation coefficient to compare the measurement data with the hypothesis. In practice, we leave it as an open problem to use more advanced tools like profiled attacks (model-less), MCP-DPA attacks \mbox{\cite{Moradi16_moments}} for the masking parts, information theoretic tools (MI/PI/HI) \mbox{\cite{gierlichs2008mutual}}, and more advanced high-dimensional attacks/filtering or post-processing.

\subsection{Attacks with PRNG off}
The PRNG generates the arithmetic shares for the adder tree, feeds the masked LUTs of the activation function and the Boolean to arithmetic converters. Turning off the PRNG unmasks all these operations making a first-order attack successful at all these points. Fig. \ref{fig:res1} shows the mean power plot on the top for orientation, which is followed below by the 3 attack plots with PRNG disabled. We attack the second stage of the adder tree, the first node's activation result,
and the first node of the output layer, respectively, shown in the next three plots. In all the plots, we observe a distinct correlation peak for the targeted variable corresponding to the correct weight and bias values.
Fig. \ref{fig:masked_dpa} shows the evolution of the correlation coefficient as the number of traces increase. The attack is successful at 200 traces with PRNG off.
This validates our claim on the vulnerability of the baseline, unprotected design.

\subsection{First-order Attacks with PRNG on}
\begin{figure*}[t!]
\vspace{-1em}
 \centering
   \includegraphics[scale=0.5]{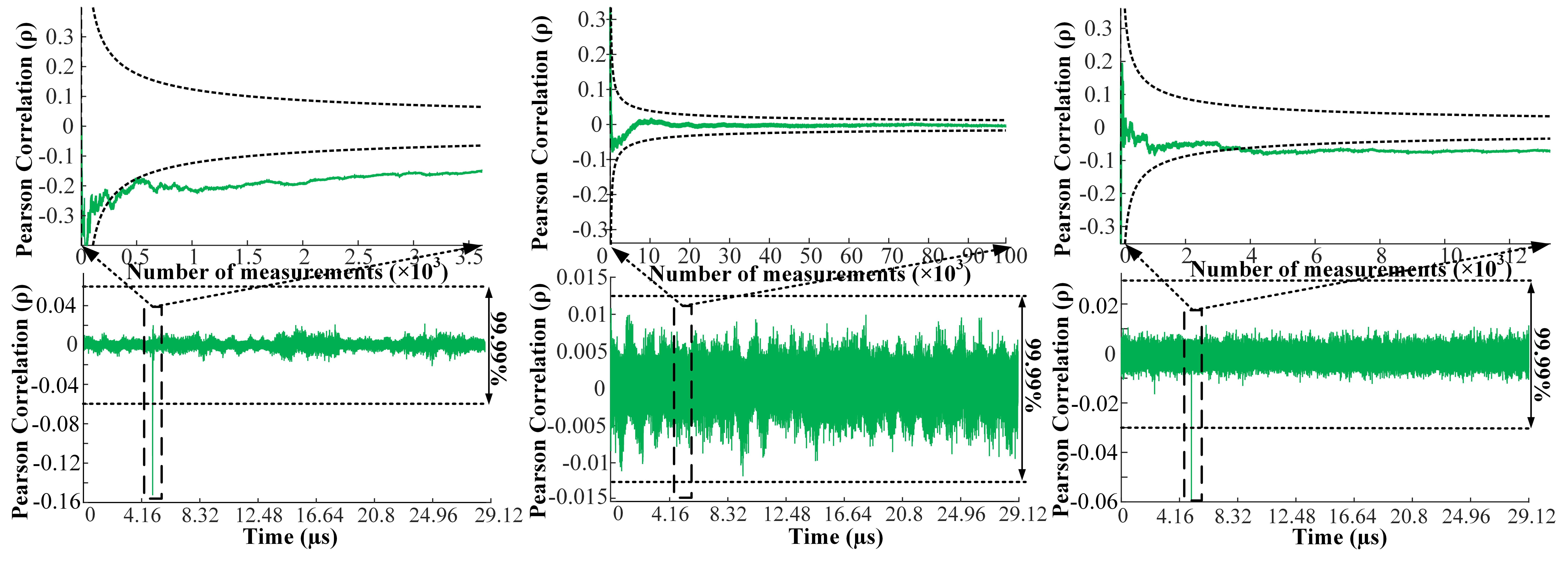}
   \vspace{-1.5em}
   \caption{Evolution of the Pearson coefficient at the point of leak with the number of traces for first-order attacks when the PRNGs are off (left), PRNGs are on (middle), and for second order attacks with PRNGs on (right). The first-order attack with PRNGs off succeeds around 200 traces but fails when PRNGs are on even with 100k traces, which shows that the design is masked successfully. The second-order attack becomes successful around 3.7k traces, which validates that we used sufficient number of traces in the first-order attacks.}
   \vspace{-1em}
 \label{fig:masked_dpa}
 \end{figure*}

We used the same attack vectors from the case of PRNG off, but with the PRNG turned on this time. This armed all the countermeasures implemented for each operation. Fig. \ref{fig:res1} shows that the distinct peaks seen in the PRNG OFF plots do not appear in the PRNG ON plots for first-order attacks.
Fig. \ref{fig:masked_dpa} shows that the first-order attack is unsuccessful with 100k traces. This validates our claim on the resiliency of the masked, protected design.

\subsection{Second Order Attacks with PRNG on}
We also performed a second-order DPA on the activation function to demonstrate that we used sufficient number of traces in the first-order attack.
Again, we used the same attack vectors used in the first-order analysis experiments, but applied the attack on centered-squared traces. Fig. \ref{fig:res1} shows that we observed a distinct correlation peak at the correct point in time.
Fig. \ref{fig:masked_dpa} shows the evolution of the correlation coefficient for the second-order attack. We can see that the attack is successful around 3.7k traces which confirms that 100k traces are sufficient for a first-order attack.

\subsection{Attacks on Hiding}
\begin{figure}[t!]
\centering
  \includegraphics[scale=0.8]{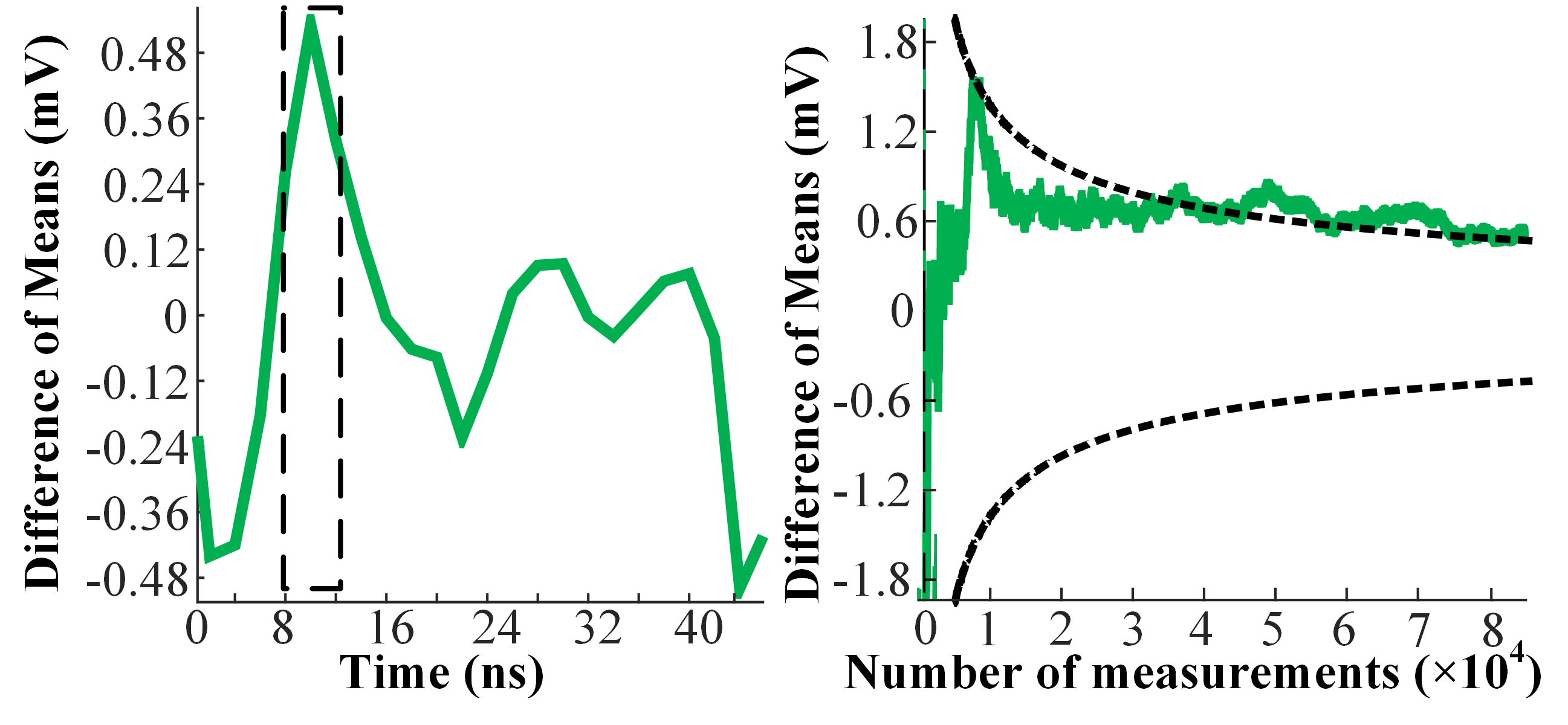} 
  \vspace{-2.5em}
  \caption{The difference-of-means test on the WDDL based signed-bit computation. The figure shows that the difference of means between the power traces corresponding to MSB=0 and MSB=1 cases cross the 99.99\% confidence interval (represented by the dotted lines) around 40k traces.}
  \label{fig:DoM}
  \vspace{-.5em}
\end{figure}

We applied a difference of means test with 100k traces to test the vulnerability of hiding used for the MSB of arithmetic shares. The partition is thus based on the binary value of MSB. Fig. \mbox{\ref{fig:DoM}} shows the attack on the targeted clock cycle and quantifies that after 40k traces the adversary is able to distinguish the two cases with 99.99\% confidence. Note that this number is significantly higher than the number of measurements required to succeed for the second order attack. The number of traces for all the attacks are relatively low due to the very friendly scenario created for the adversary; the platform is low noise. In a real-life setting, the noise would be much higher and consequently all attacks would require more traces.

\subsection{Masking Overheads}
Table \ref{tab:area} summarizes the area and latency overheads of masking in our case.
As expected, due to the sequential processing with the two shares in the masked implementation, the inference latency is approximately doubled, from $3192$ cycles for the baseline to $7248$ cycles. Table \ref{tab:area} also compares the area utilization of the unmasked vs. masked implementations in terms of the various blocks present in the FPGA. The increase in the number of the LUTs, flip flops and BRAMs in the design is approximately 2.7x, 1.7x and 1.3x. The significant increase in the number of LUTs is mainly due to the masked LUTs used to mask the activation function and convert the Boolean shares of each layer to arithmetic shares. The increase in the number of flip flops and BRAM utilization is caused by additional storage structures of the masked implementation such as the randomness buffered at the start to mask the various parts of the inference engine. Furthermore, the arithmetic masks are also stored in the first phase, to be sent together to the masked activation function later. Each layer also stores twice the number of activations in the form of two Boolean shares increasing the memory overhead.

\begin{table}[t!]
\centering
\vspace{0em}
\caption{Area and Latency Comparison of unmasked vs. masked implementations}
\label{tab:area}
\vspace{-1em}
\begin{tabular}{ |c|c|c|c| } 
\hline
 Design Type & LUT/FF & BRAM/DSP & Cycles\\ 
 \hline
 Unmasked & 20296/18733 & 81/0 & 3192\\ 
 \hline
 Masked & 55508/33290 & 111/0 & 7248\\ 
\hline
\end{tabular}
\vspace{-1.25em}
\end{table}
\section{Discussions}
This section discusses the orthogonal aspects together with the limitations of our approach and comments on how they can be complemented to improve our proposed effort.

\subsection{Limitations of The Proposed Defense}
Masking is difficult---after 20 years of AES masking, there is still an increasing number of publications (e.g. CHES'19 papers \mbox{\cite{cassiers19}, \cite{sugawara19}, \cite{DeMeyer19})} on better/more efficient masking.
This, in part, is due to ever-evolving attacks~\mbox{\cite{Levi19}}.
The paper's focus is on empirical evaluation of security.
We provide a proof-of-concept which can be extended towards building more robust and efficient solutions.
We emphasize the importance of theoretical proofs \mbox{\cite{moos2019glitch}} and the need to conduct further research on adapting them to the machine learning framework.

We have addressed the leakage in the sign bit of arithmetic share generation of the adder tree through hiding for cost-effectiveness. This is the only part in our hardware design that is not masked and hence may be vulnerable due to composability issues or implementation imbalances (especially for sophisticated EM attacks \mbox{\cite{immler2017your}}). We highlight this issue as an open problem, which may be addressed through extensions of gate level masking.
But such an implementation will incur significant overheads in addition to what we already show.

Our evaluations build on top of model-based approaches, which can be corroborated with more sophisticated attacks such as template based \mbox{\cite{chari03}}, moments-correlating based \mbox{\cite{Moradi16_moments}}, deep-learning based \mbox{\cite{maghrebi2016breaking}}, or horizontal methods \mbox{\cite{clavier2010horizontal}}. More research is needed to design efficient masking components for neural network specific computations, extending first-order masks to higher-order, and on investigating the security against such side-channel attacks.

\subsection{Comparison of Theoretical Attacks, Digital Side-Channels, and Physical Side-Channels}
We argue that a direct comparison of the physical side-channels to digital and theoretical attack's effectiveness (in terms of number of queries) is currently unfair due to \emph{immaturity} of the model extraction field and due to different target algorithms.
Analyzing and countering theoretical attacks improve drastically over time.
This has already occurred in cryptography: \emph{algorithms have won}~\mbox{\cite{kocher-talk}}.
Indeed, there has been no major breakthrough on the cryptanalysis of encryption standards widely-used today.
But side-channel attacks are still commonplace and are even of growing importance.
While digital side-channels are imminent, they are relatively easier to address in application-specific hardware accelerators/IPs that enforce constant time and constant flow behavior (as opposed to general purpose architectures that execute software). 
For example, the hardware designs we provide in this work has no digital side-channel leaks.
Physical side-channels, by contrast, are still observed in such hardware due to their data-dependent, low-level nature; and therefore require more involved mitigation techniques.

\subsection{Scaling to other Neural Networks}
The primary objective of this paper is to provide the first proof-of-concept of both power side-channel attacks and defenses of NNs in hardware. To this end, we have designed a neural network that encompasses all the basic features of a binarized neural network, like binarized weights and activations, and the commonly used sign function for non-linearity. When extended to other neural networks/datasets, like CIFAR-10, the proposed defences will roughly scale linearly with the node, layer count and bit-precision (size) of neurons. To deploy the countermeasures on constrained devices, the area overheads can be traded off for throughput, or vice versa. 
Any algorithm, independent of its complexity, can be attacked with physical side-channels.
But the attack success will depend on the parallelization level in hardware.
In a sequential design, increasing the weight size (e.g. moving from one bit to 8-bits or floating point) may even improve the attack because there is more signal to correlate.

%\vspace{-.25em}
\section{Conclusion}
%\vspace{-.25em}
\label{sec:conclusion and future work}
Physical side-channel leaks in neural networks call for a new line of side-channel analysis research because it opens up a new avenue of designing countermeasures tailored for the deep learning inference engines. 
In this paper, we provided the first effort in mitigating the side-channel leaks in neural networks. 
We primarily apply masking style techniques and demonstrate new challenges together with opportunities that originate due to the unique topological and arithmetic needs of neural networks.
Given the variety in neural networks with no existing standard and the apparent, enduring struggle for masking, there is a critical need to heavily invest in securing deep learning frameworks.

\section{Acknowledgements}
We thank the anonymous reviewers of HOST for their valuable feedback and to Itamar Levi for helpful discussions.
This project is supported in part by NSF under Grants No. 1850373 and SRC GRC Task 2908.001.

\bibliographystyle{IEEEtran}
\bibliography{paper.bib}

\end{document}